\documentclass[
superscriptaddress,
 amsmath,amssymb,
 aps,
 pre,
twocolumn
]{revtex4-2}

\usepackage{graphicx}
\usepackage{dcolumn}
\usepackage{placeins}
\usepackage{bm}
\usepackage{physics}
\usepackage{xcolor,soul}
\colorlet{darkred}{red!85!black}
\colorlet{darkgreen}{green!50!black}
\colorlet{darkblue}{blue!60!black}
\usepackage[
colorlinks   = true, 
urlcolor     = darkgreen, 
linkcolor    = darkblue, 
citecolor   = darkred 
]{hyperref}

\usepackage[normalem]{ulem}

\DeclareRobustCommand{\vect}[1]{
  \ifcat#1\relax
    \boldsymbol{#1}
  \else
    \mathbf{#1}
  \fi}

  \newcommand{\av}[1]{\left\langle#1\right\rangle}
  \newcommand{\cbr}[1]{\left(#1\right)}

  \newcommand{\wtr}{\widetilde{\rho}}
  
  \newcommand{\wtp}{\widetilde{\psi}}

\begin{document}

\title{Minimal work protocols for inertial particles in non-harmonic traps}

\author{Julia Sanders}
 \affiliation{Department of Mathematics and Statistics of the University of Helsinki, 00014 Helsinki, Finland}

\author{Marco Baldovin}
 \email{marco.baldovin@cnr.it}
\affiliation{Institute for Complex Systems, CNR, 00185, Rome, Italy}

\author{Paolo Muratore-Ginanneschi}%
\affiliation{Department of Mathematics and Statistics of the University of Helsinki, 00014 Helsinki, Finland}%

\DeclareRobustCommand{\vect}[1]{
  \ifcat#1\relax
    \boldsymbol{#1}
  \else
    \mathbf{#1}
  \fi}

\def\q{\boldsymbol q}
 \newcommand{\ti}{t_{\iota}}
 \newcommand{\tf}{t_{\styleB{f}}}


\newcommand{\styleA}[1]{\mathrm{#1}} 
\newcommand{\styleB}[1]{\mathfrak{#1}} 
\newcommand{\styleC}[1]{{#1}}  
\newcommand{\styleD}[1]{\mathscr{#1}}
 
\newcommand{\nq}{x}
\newcommand{\nt}{s}
\newcommand{\ttf}{\nt_f}
\newcommand{\variance}{v}

\newcommand{\np}{\styleA{p}}
\newcommand{\nx}{\styleA{x}}
\newcommand{\nv}{\styleA{V}}
\newcommand{\nU}{\styleA{U}}
\newcommand{\nf}{\styleA{f}}
\newcommand{\nl}{\styleA{L}}
\newcommand{\nK}{\styleA{K}}
\newcommand{\nS}{\styleA{S}}
 \newcommand{\slt}[1]{\bm{\nt}_{#1}}

 \newcommand{\bcbr}[1]{\big(#1\big)}

\newcommand{\vv}{;}

\newcommand{\syma}{\mbox{[\textbf{KL}]}}
\newcommand{\symb}{\mbox{[\textbf{EP}]}}
  
  \newcommand{\important}[1]{\textbf{#1}}
  \newcommand{\marco}[1]{\textcolor{darkgreen}{#1}}
  
\makeatletter
\let\orgdescriptionlabel\descriptionlabel
\renewcommand*{\descriptionlabel}[1]{%
	\let\orglabel\label
	\let\label\@gobble
	\phantomsection
	\edef\@currentlabel{#1}%
	\let\label\orglabel
	\orgdescriptionlabel{#1}%
}
\makeatother

\date{\today}

\begin{abstract}

Progress in miniaturized technology allows us to control physical systems at nanoscale 
with remarkable precision. 
Experimental advancements have sparked interest in
control problems in stochastic thermodynamics, typically concerning a time-dependent potential applied to a nanoparticle to reach a target stationary state in a given time with minimal energy cost. We study this problem for a particle subject to thermal fluctuations in a regime that takes into account the effects of inertia, and, building on the results of a previous work, provide a numerical method to find optimal controls even for non-Gaussian initial and final conditions, corresponding to non-harmonic confinements. The control protocol and the time-dependent position distribution are qualitatively different from the corresponding overdamped limit: in particular, a symmetry of the boundary conditions, which is preserved in the absence of inertia, turns out to be broken in the underdamped regime. We also show that the momentum mean tends to a constant value along the trajectory, except close to the boundary, while the evolution of the position mean and of the second moments is highly non-trivial. Our results also support that the lower bound on the optimal entropy production computed from the overdamped case is tight in the adiabatic limit.
\end{abstract}

\maketitle


\section{Introduction}
The past few years have seen growing interest in the experimental control of nanosystems~\cite{ MaPeGuTrCi2016, GonzalezBallestero2021, dago22, RaGuGuOdTr2023}. 
Steering the state of a nanoscale device with high precision is intriguing for many reasons, in particular for information-technology applications~\cite{lopez2016sub, Deshpande2017, ciampini2021experimental}.
For instance, a nanoparticle trapped in a bi-stable potential can model the minimal realization of an information bit: controlling nanoparticles in non-harmonic traps is therefore particularly interesting in the context of bit manipulation, specifically in the study of Landauer's erasure problem~\cite{landauer61,EsVadeBr2011,BeArPeCiDiLu2012,proesmans2020,DaPeBaCiBe2021,dago22,DaCiBe2023,DaCiBe2023b}. 
 Considerable effort has been put into finding a theoretical solution to optimal control problems for physically relevant Markov models of the dynamics: overdamped ~\cite{schmiedl2007, GoScSe2008,AuMeMG2011,AuGaMeMoMG2012, baldovin2022, guery2023driving, ChRo2023,loos2024universal}, discrete configuration space jump \cite{EsKaLiVa2010,MGMePe2012,ReSe2021,DecA2022}, and underdamped \cite{GoScSe2008,PMG2014,MGSc2014} processes.  In the underdamped case, however, besides harmonic confinements, a detailed quantitative understanding of optimality still poses an open question.

Motivated by the above, 
we search for protocols that steer a Brownian particle between two stationary end-states corresponding to different confining potentials while minimizing the average work. We consider the case of a particle subject to an underdamped dynamics characterized by small but non-negligible inertial effects. In Ref.~\cite{sanders2024optimal}, we map the problem into a system of two coupled partial differential equations, whose analytical solution is only available if the initial and final distributions of the stochastic particle are Gaussian (i.e., if the external potential fixing the boundary conditions is harmonic at the beginning and at the end of the protocol). Here, we discuss a numerical method for solving those equations for non-Gaussian boundary conditions. This gives an optimal solution to the problem for non-harmonic trapping potentials, including bi-stable potentials.

The paper is organized as follows. In Section~\ref{sec:problem} we introduce the stochastic model of interest and we formulate the optimal control problem we aim to solve.  Analytical results are known for this problem, in a suitable perturbative regime, provided that the solution to a certain differential system can be found. These analytical findings, which are not new but are needed for the following, are briefly recalled in Section~\ref{sec:analytical}. In Section~\ref{sec:numerical} we provide a numerical method to solve the mentioned differential system in the presence of generic boundary conditions: this computational tool allows us to provide an explicit solution to the problem, even when the initial and final conditions are generated by generic, non-harmonic potentials. Section~\ref{sec:results} is devoted to the discussion of the numerical results obtained in this way: qualitative differences with respect to the overdamped limit are highlighted. Finally in Section~\ref{sec:conclusions} we draw our conclusions and outline future perspectives stemming from the present work.

\section{Optimization problem}
\label{sec:problem}
Consider a classical particle of mass $m$, whose position and momentum along a given direction are denoted by $q$ and $p$ respectively. The particle is subject to a frictional force $-p/\tau$ and thermal fluctuations at inverse temperature $\beta$. We also assume the presence of a time-dependent confining potential $U(q,t)$ that can be controlled externally. The dynamics read therefore
\begin{subequations}
\label{eq:dynor}
    \begin{eqnarray}
        \dot{q}&=&\frac{p}{m}\\
        \dot{p}&=&-\frac{p}{\tau}-\partial_q U+\sqrt{\frac{2 m}{\tau \beta}}\,\xi
    \end{eqnarray}
\end{subequations}
where $\xi(t)$ is a Gaussian white noise with zero mean and $\av{\xi(t)\xi(t')}=\delta(t-t')$. At stationarity, the joint probability density function $f(q,p,t)$ of the particle's position and momentum, i.e. the solution of the Fokker-Planck equation corresponding to~\eqref{eq:dynor}, is
$$
f_{eq}(q,p|U_{eq})\propto\exp\cbr{-\beta\frac{p^2}{2m} -\beta U_{eq}(q)}\,.
$$
Controlling $U(q,t)$ allows us to modify the distribution of the system in time. We want to steer the state from an initial probability density function 
$$
f_{\iota}(q,p)=f_{eq}(q,p|U_{\iota})
$$
to a final one 
$$
f_{f}(q,p)=f_{eq}(q,p|U_{f})
$$
in a finite time $t_f$ while minimizing the average external work done on the system
\begin{equation}\label{eq:cost_functional}
    \mathcal{W}=\int_0^{t_f}dt \int dq\, dp\, \partial_t U(q,t) f(q,p,t)\,.
\end{equation}
Minimizing the average work between fixed states is equivalent to finding the protocol leading to the minimal average entropy production within this framework~\cite{AuGaMeMoMG2012,PePi2020}. The cost of the corresponding optimization problem is, however, non-convex in the control~\cite{GoScSe2008,MGSc2014}.
Following the strategy of~\cite{MGSc2014}, it is therefore useful to consider a regularized version of the dynamics, characterized by the same stationary distribution:
\begin{subequations}
\label{eq:dyn}
    \begin{eqnarray}
        \dot{q}&=&\frac{p}{m}-\frac{g \tau}{m}\,\partial_q U +\sqrt{\frac{2g\tau}{m \beta}}\,\eta \label{eq:dyna}\\
        \dot{p}&=&-\frac{p}{\tau}-\partial_q U+\sqrt{\frac{2 m}{\tau \beta}}\,\xi
    \end{eqnarray}
\end{subequations}
where $\eta(t)$ is a Gaussian white noise independent of $\xi(t)$, such that $\av{\eta(t)\eta(t')}=\delta(t-t')$. Here $g$ is a regularizing dimensionless constant: in the limit $g \to 0$, the dynamics~\eqref{eq:dynor} are recovered. A physical interpretation of the above dynamics comes from considering an additional thermal bath acting on the particle, with damping coefficient $(g \tau)^{-1}$. If $g$ is small enough, it is legitimate to approximate the effect of this additional bath as an overdamped contribution to the particle's velocity, providing the correction in Eq.~\eqref{eq:dyna}.
The cost function of the regularized process, i.e. the average entropy production, is given by~\cite{sanders2024optimal}
   \begin{equation}
\label{model:ep}
\begin{aligned}
\mathcal{E}=&\av{\ln\dfrac{f_{\iota}(q,p)}{f_{f}(q,p)}}+\frac{1}{\tau}\int_0^{t_f}dt\,\cbr{\frac{\beta \av{p^2(t)}}{m}-1}\\
&+\dfrac{g\,\tau}{m}\int_{0}^{t_f}dt\,\av{\beta\cbr{\partial_q U(t)}^2-\partial_q^2 U(t)}.    
\end{aligned}
	\end{equation} 
The first term is the Gibbs-Shannon entropy variation between the end states, which is fixed by the boundary conditions. In the $g\ll 1$ limit the relevant part of the cost comes from the second term, which forces the momentum variance to stay as close as possible to its equilibrium value.

We address the problem within the mathematical framework of Pontryagin's theory~\cite{kirk2004optimal,BecJ2021}. We need to minimize the action
\begin{align}
	\hspace{-0.3cm}	\mathcal{A}[\styleC{f},U,V]&=\mathcal{E}+\av{V(q,p,0)}_{\iota}
	-\av{V(q,p,t_f)}_f
	\nonumber\\
	&+\int_{0}^{t_f}\,dt\,\av{(\partial_{t}+\styleB{L})V(q,p,t)}\,,
	\label{oc:BP}
\end{align} 
where $V$ is a Lagrange multiplier enforcing the dynamics (it is equivalent to the value function of the Hamilton-Jacobi-Bellman formalism). The operator
\begin{equation}
\label{eq:FPoperator}
\begin{aligned}
\styleB{L}&=	\dfrac{p-\tau\,g\,(\partial_q U)}{m}\cdot\partial_{q}
-\left(\dfrac{p}{\tau}+\partial_q U\right)\cdot\partial_{p}\\
&+\dfrac{g\,\tau}{m\,\beta}\,\partial_{q}^{2}
+\dfrac{m}{\tau\,\beta}\,\partial_{p}^{2}\,
\end{aligned}    
\end{equation}
is the one associated to the backward Fokker-Planck dynamics corresponding to the evolution~\eqref{eq:dyn}.
The averages $\av{\cdot}_{\iota}$ and $\av{\cdot}_f$ are computed over the boundary distributions $f_{\iota}(q,p)$ and $f_f(q,p)$.

\section{Summary of analytical results}
\label{sec:analytical}

An analytical study of the problem in the previous section was carried out by the authors of this paper in~\cite{sanders2024optimal}. It is shown there that in order to find the optimal control, one first needs to solve the ``cell problem'' in dimensionless variables
\begin{subequations}
\label{eq:cell}
    \begin{eqnarray} \label{eq:cellA}
			\partial_{\nt}\rho &=&  \varepsilon^{2}\partial_{\nq}\rho\,\left(\partial_{\nq} \sigma \right)\\
	\label{eq:cellB}\partial_{\nt}\sigma &=&\dfrac{\varepsilon^2}{2}(\partial_{\nq} \sigma)^2
		\end{eqnarray}
\end{subequations}
for the (dimensionless) fields $\rho(\nq,\nt)$ and $\sigma(\nq,\nt)$, with boundary conditions given by
\begin{subequations}
\label{eq:boundary_conditions}
\begin{align}\label{eq:boundary_conditions_initial}
\rho(\nq,0)&=\frac{\exp\cbr{-\beta U_{\iota}(\ell \nq)}}{\int_{\mathbb{R}}dy\, \exp\cbr{-\beta U_{\iota}(y)}}\\
\label{eq:boundary_conditions_final}\rho(\nq,t_f/\tau)&=\frac{\exp\cbr{-\beta U_{f}(\ell \nq)}}{\int_{\mathbb{R}}dy\, \exp\cbr{-\beta U_{f}(y)}} \,.  
\end{align}    
\end{subequations}
The parameter $s$ can be interpreted as a dimensionless time and $x$ as a dimensionless space. We stress that the above boundary conditions fully specify the problem, and there is no need for additional conditions on $\sigma$. The dimensionless parameter
\begin{equation}
    \varepsilon=\dfrac{\tau}{\sqrt{\beta \, \ell^2\, m}}\,,
\end{equation}
where $\ell$ represents a characteristic length-scale of the model~\footnote{E.g., $\ell=\sqrt{\av{q^2}_{\iota}-\av{q}_{\iota}^2}\,$,
where the averages are computed with respect to the initial state.}, is assumed to be much smaller than 1. The value of $\varepsilon$ sets the time-scale separation between the thermal relaxation of the momenta and the dynamical evolution, with the limit $\varepsilon \to 0$ corresponding to an overdamped regime where the momenta relax instantaneously.
It can be shown that the field $\rho$ solving Eq.~\eqref{eq:cell} is related to the solution of the optimal problem for an overdamped dynamics with the same boundary conditions~\cite{AuGaMeMoMG2012}.
It turns out that a perturbative solution to the optimal underdamped problem can be found up to second order in powers of $\varepsilon$
 for a time horizon $t_f\simeq \tau O(\varepsilon^{-2})$, once $\rho$ and $\sigma$ are known.  Let us notice that this time interval diverges as $\varepsilon^{-2}$ if we take the limit $\varepsilon \to 0$ keeping $\tau$ fixed, while it diverges as $\varepsilon^{-1}$ if the typical mechanical frequency is kept constant and $\tau$ is sent to zero. In both cases, this scenario corresponds to a long driving-time limit, with respect to the characteristic times of the system.
 The main purpose of this paper is to provide a numerical method of computing the solution of the cell problem~\eqref{eq:cell}, which in general cannot be achieved with analytical techniques.

For the sake of self-consistency, we summarize here the main findings of~\cite{sanders2024optimal}, which show how the solution of the underdamped problem is found, once the solution of~\eqref{eq:cell} is known. 
We refer the interested reader to the original work for details on the derivation, and we provide instructions in Appendix~\ref{sec:appnotation} on switching between the two notations. 
It is useful to define two auxiliary functions
	\begin{subequations}
		\begin{align}
a({\nt})&=1+\sinh (\omega \nt) \tanh \frac{ \omega \ttf}{2}-\cosh ( \omega \nt)+b({\nt})
\\ 
b({\nt})&=\frac{\omega\,e^{-\ttf}\,
	\left(\cosh (\omega\nt)-e^{2 \nt}\right)}{\omega  \cosh \ttf-2 \sinh  \ttf \coth \frac{ \omega \ttf}{2}}	\\
	&\hspace{0.5cm}	+		\frac{\omega\,e^{-\ttf} \left(e^{2 \ttf}-\cosh (\omega \ttf)\right)\, \sinh ( \omega\nt )
	}{\left(\omega  \cosh \ttf-2 \sinh  \ttf \coth \frac{ \omega \ttf}{2}\right)\sinh( \omega\ttf)}\,,	\nonumber	
	\end{align}
	\end{subequations}
 where $\ttf=t_f/\tau$ and
$$
\omega=\sqrt{\dfrac{1+g}{g}}\,.
$$
Let us denote their integral averages in $[0,\ttf]$, rescaled by $(1+g)^{-1}$, as
\begin{subequations}
    \begin{align}
        A&=\frac{1+g}{s_f}\int_0^{\ttf}d\nt \, a(\nt)\,\\
        B&=\frac{1+g}{s_f}\int_0^{\ttf}d\nt \, b(\nt)\,.
    \end{align}
\end{subequations}
 It is also useful to introduce the mean and variance of the distribution $\rho$, namely
\begin{subequations}
\begin{eqnarray}
    \mu(s)&=&\int_{\mathbb{R}}d\nq\, \rho(\nq,s) \nq \label{eq:mu}\\   
    \variance(s)&=&\int_{\mathbb{R}}d\nq\, \rho(\nq,s) \nq^2-\mu(s)^2\,.
\end{eqnarray}    
\end{subequations}
It can be proven that
$$
\mu'(s)=\frac{\tau}{t_f}\frac{\av{q}_f-\av{q}_{\iota}}{\ell}\,,
$$
where the averages $\av{\cdot}_{\iota}$ and $\av{\cdot}_{f}$ are computed over the initial and final state, respectively: this is briefly recalled in Appendix~\ref{app:symm}.
Equipped with the above definitions, we can write the explicit form of the control potential providing the optimal solution, namely
\begin{equation}
\label{eq:control}
\begin{aligned}
	     U(q,t)\simeq &-\frac{1}{\beta}\ln\rho(q/\ell,t/\tau)
		 +c_1(t/\tau) \sigma(q/\ell,t/\tau)\\
   &-c_2(t/\tau)  q   +O(\varepsilon)    
\end{aligned}
\end{equation}
with
\begin{subequations} 
\begin{eqnarray}
c_1&=& \frac{1}{\beta}\frac{a'+a}{A} \\
  c_2&=&\frac{m \ell }{\tau^2}\frac{B\,a'-A \,b'+B\,a-A \,b}{ A\,(A-B)}\mu'\,\,.  
\end{eqnarray} 
\end{subequations}
Here, the prime denotes the derivative of the function with respect to its argument: the dimensionless time. 
  When not specified, the functions $a$, $b$, $\mu$, $v$ and their derivatives depend on the rescaled time $t/\tau$: we drop the explicit dependence to lighten the notation. We recall that, in the overdamped dynamics, the optimal control potential reads~\cite{AuMeMG2011}
 \begin{equation}
 \label{eq:uover}
     U_{over}(q,t)=\frac{1}{\beta}\cbr{ \sigma(q/\ell,t/\tau) -\log(\rho(q/\ell,t/\tau))}\,.
 \end{equation}

We can also characterize the evolution of the system during the process. The marginal probability density function reads
\begin{equation}
\label{eq:f}
    f(q,t)\simeq \frac{\rho(q/\ell, t/\tau)}{\ell}+\frac{\varepsilon^2}{\ell}f_2(q/\ell, t/\tau)+O(\varepsilon^3)\,,
\end{equation}
where the leading correction is given by
\begin{equation}  \label{eq:correction_to_density}  
		\begin{aligned}
			f_2&(x,t)=-g \,\partial_{\nq} f_1(x,t/\tau,t/\tau) \\
   &+(1+g) \left(\dfrac{t}{ \tf}\int_{0}^{\tf/\tau}\mathrm{d}s  -\int_{0}^{t/\tau}\mathrm{d}s \right)\partial_{\nq}f_1(x,s, t/\tau)\,
		\end{aligned}
\end{equation}
with
\begin{equation}
\begin{aligned}
    f_1(\nq,s,s')=&-\frac{a(s)}{A} \rho(\nq,s')\partial_{\nq} \sigma (\nq,s')\\
    &+\frac{Ba(s)-Ab(s)}{\varepsilon^2 A(A-B)}\rho(\nq,s') \mu'(s') \,.
\end{aligned}			
\end{equation}
In the overdamped limit, one has
\begin{equation}
\label{eq:fover}
    f_{over}(q,t)=\frac{\rho(q/\ell,t/\tau)}{\ell}\,,
\end{equation}
consistent with~\eqref{eq:f}.

The first moments are given by 
\begin{equation}\label{eq:mommean}
   \av{p(t)} 
     \simeq  \frac{m \,\ell}{\tau}\,\frac{a-b}{A-B}\,\mu'+O(\varepsilon^{2}) 
\end{equation}
and
 \begin{equation}
 \begin{aligned}\label{eq:posmean}
\av{q(t)}\simeq &\,\ell \cbr{\mu-\frac{t}{\tau}\mu'}	+	\frac{g \,\tau}{m} \, \av{p(t)} \\
&+\frac{\ell\,(1+g)\,\mu'}{A-B}  \int_{0}^{t/\tau}\mathrm{d}s\,   \bcbr{a({s})-b({s})}+O(\varepsilon^{3})  \,.   
 \end{aligned}
 \end{equation}
The second moments read
\begin{equation}
    \begin{aligned}\label{eq:posvar}
 \av{q(t)^2}&-\av{q(t)}^{2}\simeq
		\ell^2\, v
		+\,\frac{g \,\ell^2}{A}\,a v'-\frac{t\,\ell^2}{\tau}\,v'\\
  &	+\frac{1+g}{A}\,\ell^2\,v'\int_0^{t/\tau}\mathrm{d}s\, a(s) +O(\varepsilon^{3})\,,
    \end{aligned}
\end{equation}
\begin{equation}
    \begin{aligned}        \label{eq:momvar}
			\av{p(t)^{2}}&- \av{p(t)}^{2}\simeq
		\frac{m }{\beta}- \frac{m^2 \,\ell^2 \,a^{2}\,(\mu')^{2}}{\tau^2 A^{2}}\\
 &+\frac{a^{2}}{A^{2}}
 \frac{\tau^2}{\beta^2\,\ell^2}
 \int_{\mathbb{R}}\mathrm{d}\nq\,
	\rho(\nq,t/\tau)\,\big{(}\partial_{\nq}
	\sigma(\nq,t/\tau)\big{)}^{2}
		\\
  			&+\frac{2}{A}
     \frac{\tau^2}{\beta^2\,\ell^2}
     \int_0^{t/\tau}ds\, e^{-2(t/\tau-s)}\,a(s)\times\\
&\times\int_{\mathbb{R}}\mathrm{d}\nq\,\rho(\nq,t/\tau)\, 
	\partial_{\nq}^2\sigma(\nq,t/\tau)+O(\varepsilon^{3})
    \end{aligned}
\end{equation}
and

\begin{equation}
  \av{q(t)p(t)}-\av{q(t)}\av{p(t)}\simeq
		 \, \frac{ \tau} {\beta}\frac{a}{2\,A}  v'
		+O(\varepsilon^{2})\,.
\end{equation}

The above results also allow us to write an explicit lower bound for the average entropy production ~\cite{sanders2024optimal}:
\begin{equation}\begin{split}
\label{eq:overdamped_lower_bound}
    \mathcal{E} &\geq \dfrac{\varepsilon^{2}}{1+g}\int_{0}^{t_f/\tau} \dd s \int_{\mathbb{R}} \dd x\, \rho(x,s) (\partial_x \sigma (x,s))^2 \,,
\end{split}\end{equation}
where the lower bound is the average entropy production for the overdamped dynamics. 

\begin{figure}\hspace*{-0.4cm}
    \includegraphics[width=1.1\linewidth]{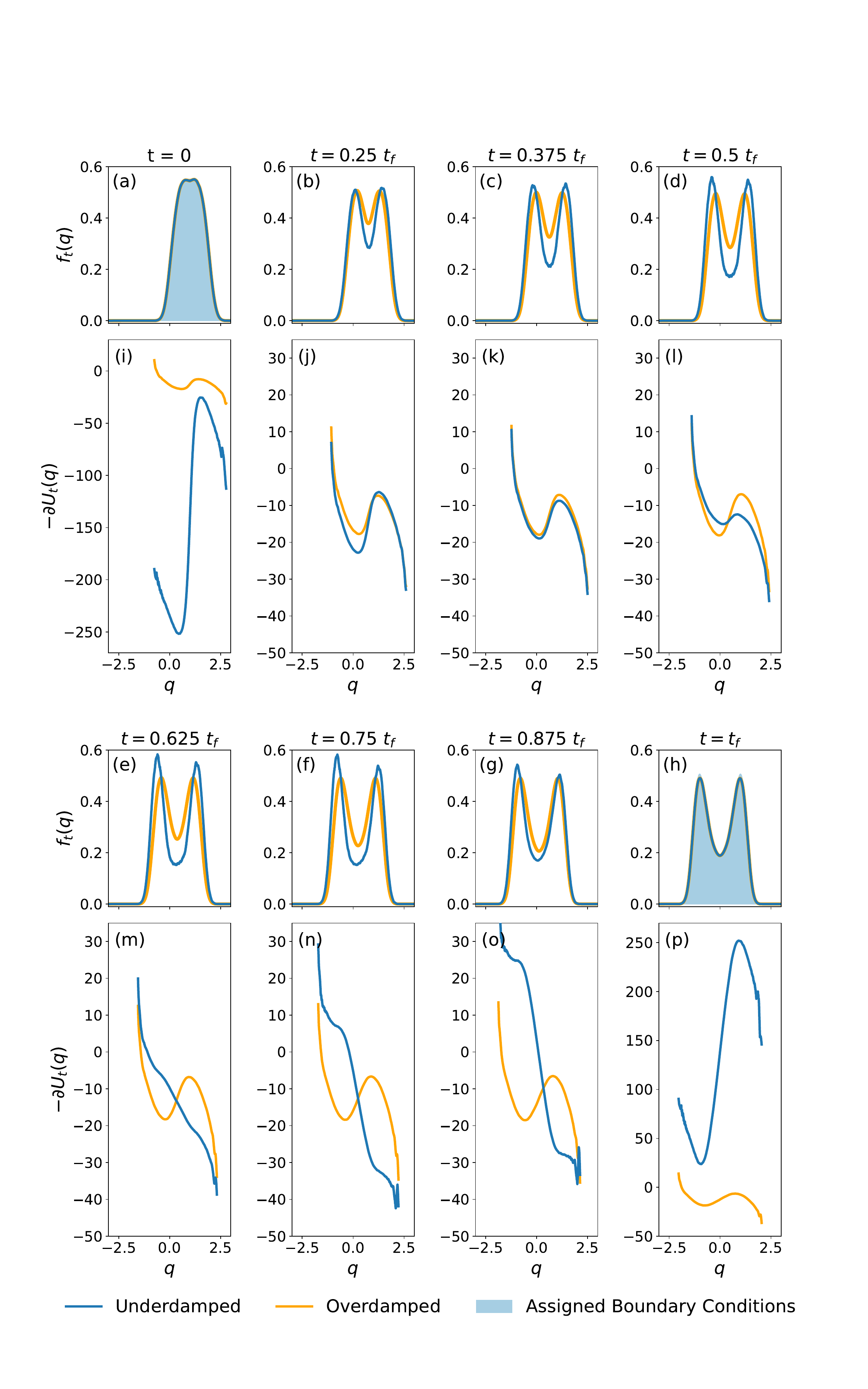}
    \caption{Optimal nucleation process. We compute the optimal evolution with boundary conditions~\eqref{eq:nucleation}, in the overdamped (orange) and underdamped (blue) case, in finite time. Panels (a)-(h) show the probability density~\eqref{eq:fover} in the overdamped dynamics, computed with the method outlined in Section~\ref{sec:numerical}, and the position marginal distribution~\eqref{eq:f} in the underdamped dynamics. Panels (i)-(p) show the drift in the two cases, computed by Eqs.~\eqref{eq:uover} and~\eqref{eq:control}. 
    Parameters of the dynamics:  $\mathrm{t}_f=2$, $\varepsilon=0.2$, $\tau=\beta=1$, and $g=0.01$. For the numerical method we use $N=2 \times 10^5$ and  $h=0.2$. The drifts are shown for area of mass greater than $10^{-4}$.}
    \label{fig:potentials}
\end{figure}

\section{Numerical method} 
\label{sec:numerical}

Once the solution of the cell problem~\eqref{eq:cell} is known, the optimal problem in the underdamped dynamics is solved up to second order in $\varepsilon$. Unfortunately, the differential system~\eqref{eq:cell} turns out to be analytically solvable only in a few special cases (notably, when the confining potentials at the boundaries, $U_{\iota}$ and $U_{f}$, are harmonic). We therefore aim at providing a numerical method to compute the solution in the presence of generic boundary conditions.

Following~\cite{AuMeMG2011} and Section V in~\cite{AuGaMeMoMG2012}, we recall that a velocity solving Burger's equation of the form \eqref{eq:cellB} is constant along trajectories of the dynamics, and therefore trajectories under this velocity must travel along straight lines. Furthermore, the initial and final positions of trajectories minimizing the entropy production solve an optimal transport problem. We can therefore linearly interpolate between paired endpoints in order to find the fields $\sigma$ and $\rho$. Our strategy is the following. First, $N$ positions are sampled independently from the initial and final distributions, denoted $\{x_{\iota}^{(n)}\}_n$ and $\{x_f^{(n)}\}_n$ respectively, for $1 \le n\le N$. The labels are chosen in such a way that, if $n<m$, $x_{\iota}^{(n)}\leq x_{\iota}^{(m)}$ and $x_f^{(n)}\leq x_f^{(m)}$. A matching is then found that minimizes the total squared distance between the two sets, preserving the ordering of points: i.e., $x_{\iota}^{(n)}$ is matched with $x_{f}^{(n)}$.  In the one dimensional case shown here, we use a linear programming 
solver from the Python Optimal Transport library~\cite{flamary2021pot} to pair the points $x_{\iota}^{(n)}$ and $x_{f}^{(n)}$ from the sampled histograms. 
Matched end-points $x_{\iota}^{(n)}$ and $x_{f}^{(n)}$ are then used to obtain a discrete approximation of the Lagrangian map at any time $t$ through a linear interpolation
\begin{equation}
\label{eq:lagrange_map}
    x^{(n)}(t)= \dfrac{t_f- t}{t_f}\,x^{(n)}_{\iota} + \dfrac{t}{t_f}\,x^{(n)}_{f}\,.
\end{equation} 
The interpolation is performed for fixed times $t$ in a discretization of $[0,t_f]$. At each time, the resulting set of points are samples from the intermediate marginal density of the position, and can be used to estimate the density $\rho$. We use kernel density estimation: a method of estimating distributions from an empirical set of samples and a smoothing parameter $h$ (the bandwidth)~\cite{ElemStatLearn}. 

An Epanechnikov kernel  
\begin{equation}
    \label{eq:epanechnikov}K(x)=\begin{cases}
        \frac{3}{4}(1-x^2) &\text{ if } |x|\leq1\,, \\
        0 &\text{ otherwise}
    \end{cases}
\end{equation} 
is used~\cite{scikit-learn}.
We then find the value of $\partial_x\sigma$. This is 
approximated by 
\begin{equation}
    \label{eq:burgersvel}-\partial_x \sigma(x^{(n)}(t),\,t/\tau) = \dfrac{\tau}{t_f} \left(x^{(n)}_{f} - x^{(n)}_{\iota}\right)\,,
\end{equation}
and then interpolated into a polynomial function. The code implementing the procedure can be found at~\cite{github}. 

The resulting fields $\rho$ and $\sigma$ are solutions of the cell problem~\eqref{eq:cell}. To find the corrections for the underdamped dynamics, we implement the expressions detailed in Section~\ref{sec:analytical}.

\section{Results}
\label{sec:results}
We discuss here a nucleation process with minimum entropy production. This means that the assigned initial distribution is a single peaked distribution, and the assigned final distribution has two peaks, representing the destruction of one bit of memory~\cite{LeOrPoSn2019}. In particular, we use the boundary conditions~\eqref{eq:boundary_conditions} with
\begin{equation}
\label{eq:nucleation}
    \begin{aligned}
        U_{\iota}(q) &= (q-1)^4\\
        U_{f}(q) &= (q^2-1)^2\,.
    \end{aligned}
\end{equation}
Figure~\ref{fig:potentials} illustrates the process, with the predicted marginal densities of the position and the drift of \eqref{eq:dyn} needed to steer the system between the assigned end points. The corresponding overdamped dynamics is also plotted for comparison. Even for the value $\varepsilon=0.2$ considered here, several qualitative differences emerge. In the overdamped limit the distribution is always symmetric with respect to its center of mass, i.e. it is invariant under $q-\av{q} \leftrightarrow \av{q}-q$: this symmetry, present in the boundary conditions that we have chosen, is indeed preserved by the optimal overdamped evolution, as briefly discussed in Appendix~\ref{app:symm}. The underdamped dynamics appears  instead to break this symmetry during the evolution. 
 We notice, in particular, that the heights of the peaks of the marginal distirbution of the position in the underdamped dynamics follow different patterns than those in the overdamped: in the former, they are equal to each other during the evolution,
 while in the latter they differ. In this sense, the difference of peak heights in the optimal overdamped dynamics appears as a signature of inertial effects.

\begin{figure}
    \centering
    \includegraphics[width=0.99\linewidth]{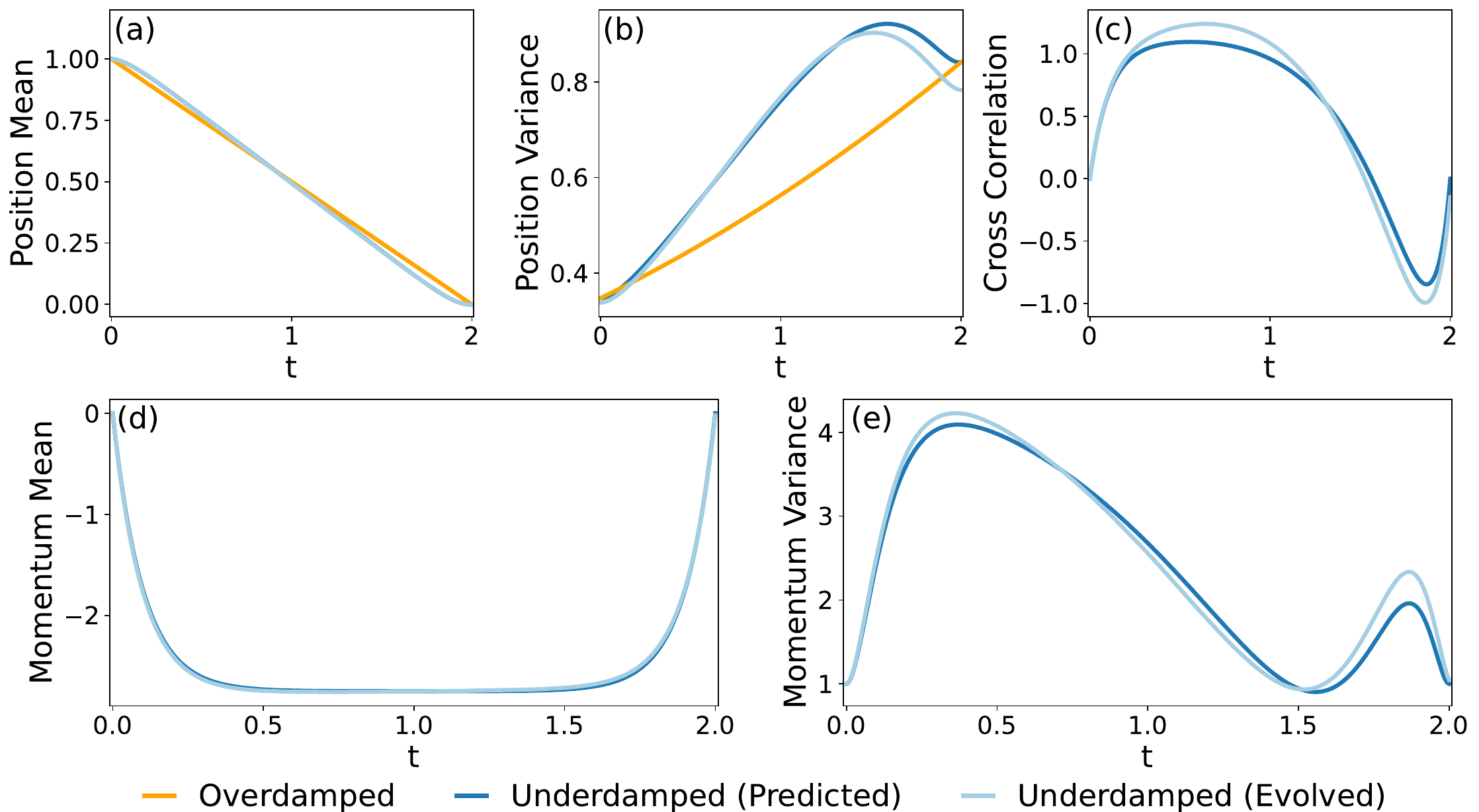}
    \caption{\label{fig:ep_land_mom_cumulants} Evolution of the cumulants. For the process of Fig.~\ref{fig:potentials}, we compute the overdamped (orange) and the underdamped moments (dark blue). We also simulate the dynamics~\eqref{eq:dyn} under the predicted underdamped optimal control protocol~\eqref{eq:control} and compute the corresponding empirical cumulants, shown in light blue. Dynamics~\eqref{eq:dyn} is simulated through a Euler-Maruyama scheme with time-step $\Delta t=0.005$, and $5 \times 10^5$ independent trajectories are considered. Boundary conditions are applied on the drift for areas of density smaller than $10^{-5}$: on the left, we use the absolute value of the boundary value and its negative on the right.}
    \label{fig:cumulants}
\end{figure}

We also remark that the shape of the optimal drift, which evolves quite smoothly in the overdamped limit, experiences instead a non-trivial dynamics in the underdamped regime, stemming from a qualitatively different control mechanism. In the absence of inertia, the computed drift is (almost) always negative, meaning that the system experiences an external force pointing in the negative direction. When inertia is considered, there is a first phase (panels (i)-(m)) in which the particle is pushed toward the negative direction, and a second one (panels (n)-(o)) where particles on the left are pushed to the right, and particles on the right are pushed to the left, in a confining-like fashion. This behaviour is even more evident in Fig.~\ref{fig:ep_land_mom_cumulants}, where we plot numerical predictions of the first and second order cumulants of the position and momentum in the underdamped dynamics~\eqref{eq:dyn}. 
In particular, the position variance evolution in Fig.~\ref{fig:ep_land_mom_cumulants}(b) shows that the optimal underdamped dynamics is obtained by first letting the position distribution spread, much faster than in the overdamped case, and then compressing it again to reach the target final state. This is consistent with the qualitative analysis of the drift proposed above.

To demonstrate self-consistency of the numerical method, the predictions for the cumulants are compared to numerical estimates of the moments computed along simulated trajectories of the underdamped dynamics under the predicted optimal drift~\eqref{eq:control}. While some deviation is to be expected within a perturbation theory framework, we note that two methods of computing the cumulants give consistent results.
The numerical predictions for the momentum cumulants display as well a behaviour characteristic to the effects of inertia. 
This is further illustrated in Fig.~\ref{fig:ep_momcumulants} for decreasing $g$. From there it is quite evident that, in the physical limit $g\to 0$, the momentum mean tends to a constant value.

\begin{figure}
\centering    
\includegraphics[width=0.99\linewidth]{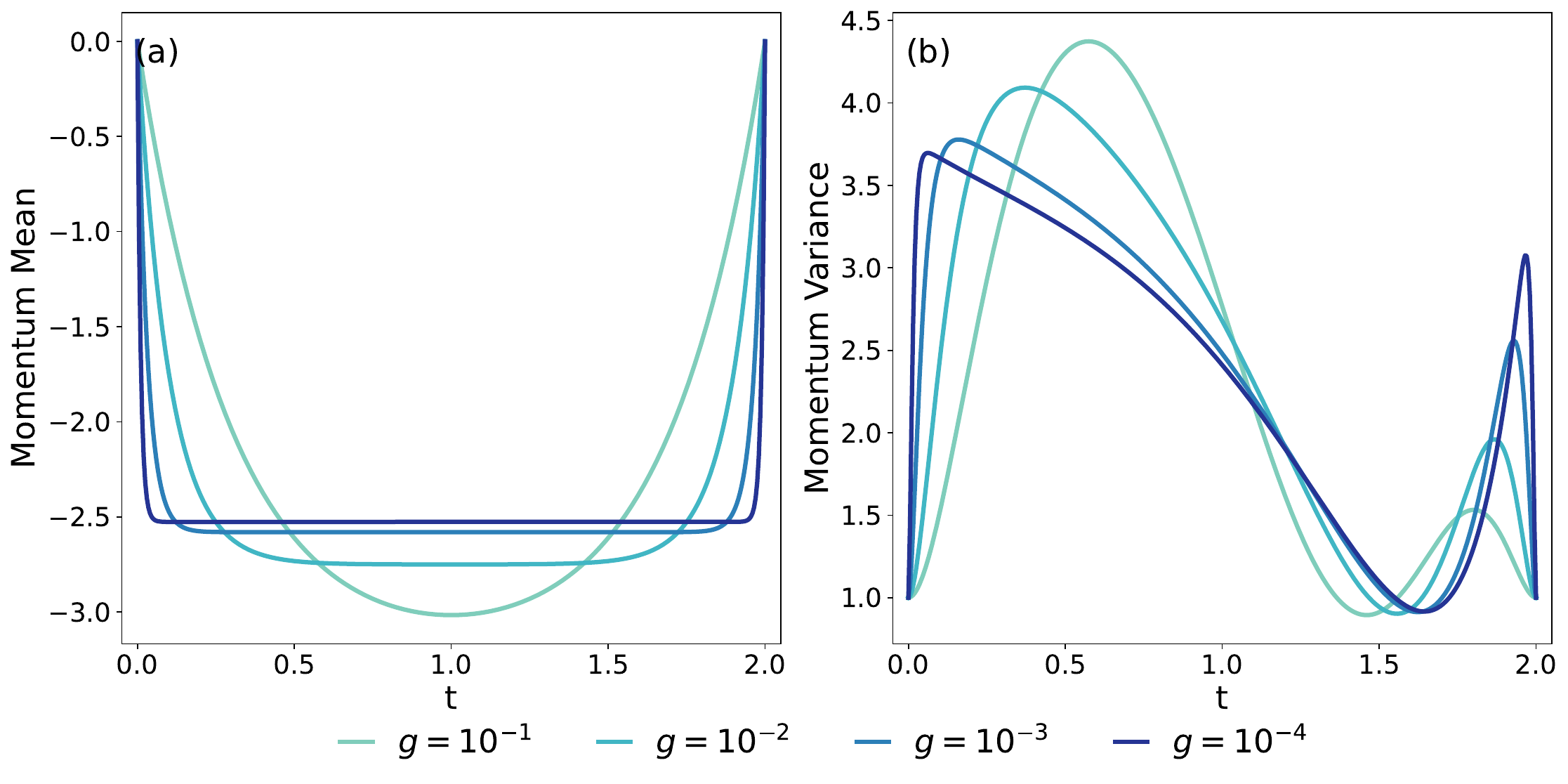}
    \caption{\label{fig:ep_momcumulants} Dependence on $g$. Momentum mean (a) and variance (b) for different values of $g$ in the nucelation process, as  functions of time. Assigned boundary conditions and all other numerical parameters and method used are the same as in Fig.~\ref{fig:potentials}}
\end{figure}

\begin{figure}
    \centering
    \includegraphics[width=0.99\linewidth]{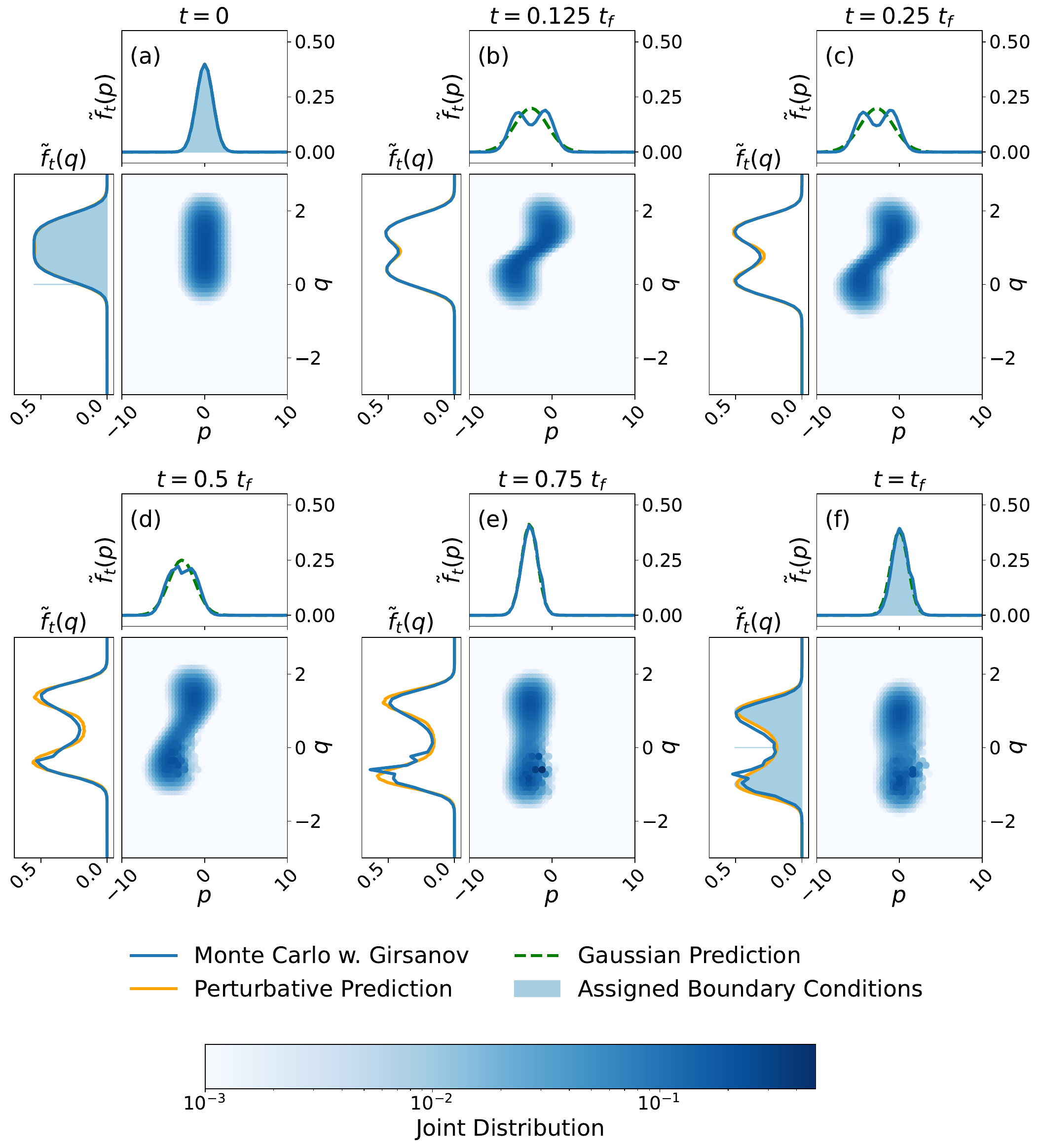}
    \caption{Solution in $(p,q)$ space.
    In each panel, we plot the joint distribution (center), and the marginal distribution of the position (left) $\tilde{f}_t(q)$ and momentum  $\tilde{f}_t(p)$ (top). Different panels correspond to different times. The joint density is computed as a numerical average of weighted sample trajectories of~\eqref{eq:dyn}. For details of the method, see Section 3 of \cite{sanders2024numericalintegration}. Trajectories are discretized by the Euler-Maruyama scheme with time step $\Delta t=0.005$ and $5\times 10^4$ independent realizations from $2.6 \times 10^3$ equally spaced points in $[-3,3]\times[-10,10]$.   
    On the left inset, the perturbative prediction for marginal density of the position in the underdamped dynamics is shown in orange, and on the top inset, the Gaussian prediction for the marginal density of the momentum is shown by the dashed green line. This is estimated by a Gaussian distribution whose moments are the momentum mean and variance as shown in Fig.~\ref{fig:cumulants}. 
    Other parameters as in Fig.~\ref{fig:potentials}.}
    \label{fig:jointdist}
\end{figure}
Fig.~\ref{fig:jointdist} shows the solution of the Fokker-Planck equation~\eqref{eq:FPoperator} driven by the perturbatively estimated drift, giving the intermediate joint density of the position and momentum. We compute this using a Girsanov weighting applied to trajectories of an associated divergence-free stochastic process; see the method described in Section 3 of~\cite{sanders2024numericalintegration}. The joint distribution allows us to compute the intermediate marginal distributions of the momenta, where we see the effects of perturbation theory. In particular, we notice the double peaked distributions shown in the top inset of panels (b), (c) and (d) of Fig.~\ref{fig:jointdist}. This most likely arises because of the lack of higher order terms in the perturbatively estimated drift $\partial U$. However, we notice that even at this order, we have good agreement with the boundary conditions in panel (f).  

\begin{figure}
\centering    
\includegraphics[width=0.99\linewidth]{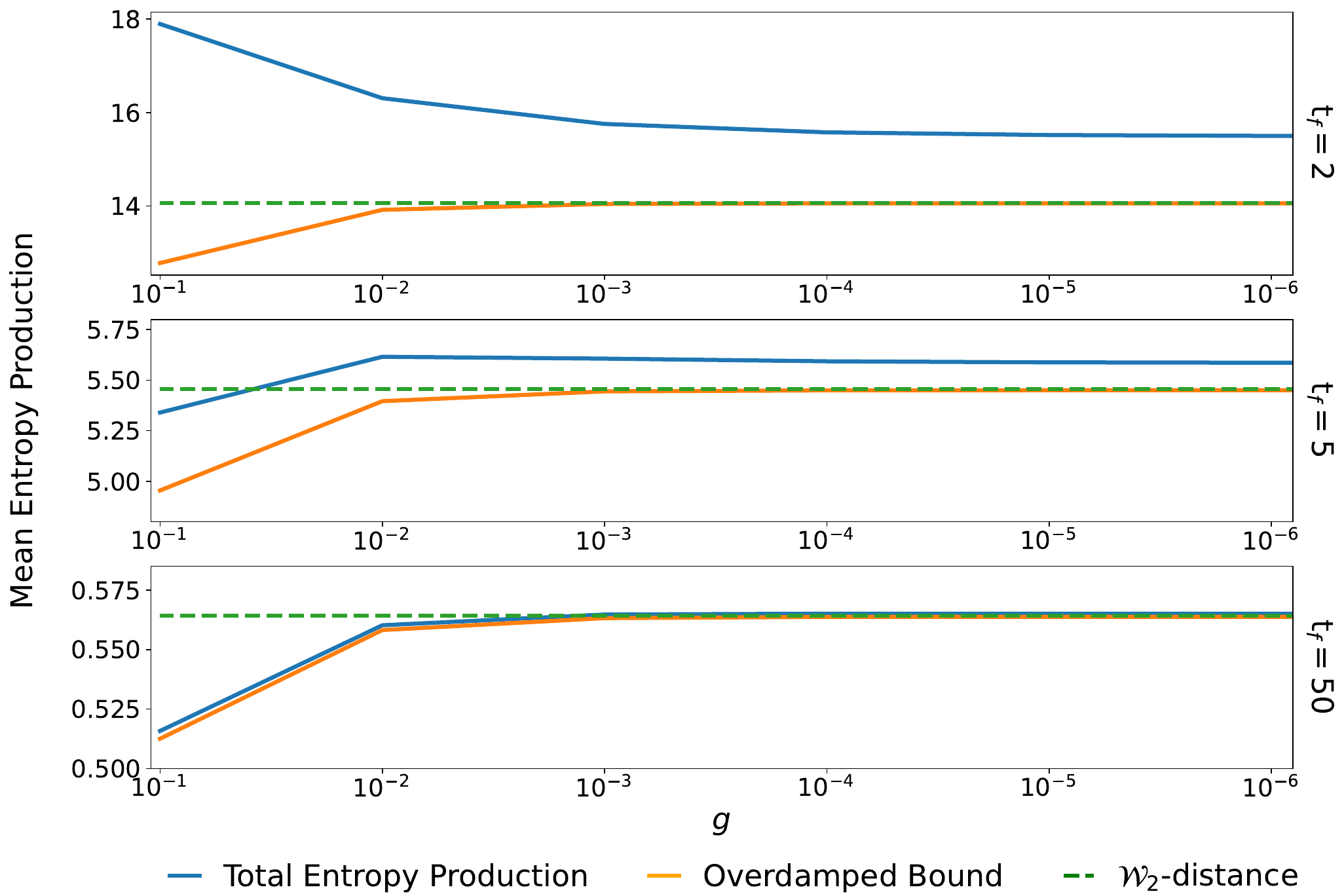}
    \caption{\label{fig:ep_cost_all} Lower bound on the cost. The total entropy production~\eqref{model:ep} (blue) of the nucleation process in the underdamped dynamics is shown for different values of $t_f$, as a function of $g$ (decreasing along the horizontal axis), and compared with the lower bound~\eqref{eq:overdamped_lower_bound} (orange). We also report, for comparison, the scaled $\mathcal{W}_2$-distance~\eqref{eq:w2_dist} (green dashed), computed from a numerical estimate of the squared Wasserstein-2 distance between the sampled histograms of initial and final distributions. At $g = 10^{-6}$, the difference between the entropy production cost and the lower bound is $1.4$ for $t_f=2$, $ 0.14$ for $t_f=5$, and  $1.3 \times 10^{-3}$ for $t_f=50$. Assigned boundary conditions and all other numerical parameters are as in Fig.~\ref{fig:potentials}. }
\end{figure}

We compute the total entropy production cost and the lower bound given by the overdamped dynamics at different values of final time $t_f$ in Fig.~\ref{fig:ep_cost_all}. This supports that the bound is tight in the adiabatic limit, as $t_f\to\infty$. 
To once again verify the consistency of the numerics, we also plot the squared Wasserstein-2 distance between the assigned initial $P_{\iota}$ and final $P_f$ distributions~\eqref{eq:boundary_conditions}
\begin{equation}\label{eq:w2_dist}
    \mathcal{W}^2_2(P_{\iota},P_f) = \inf_{\pi}\int \dd \pi(x,y)\left(x-y\right)^2\,.
\end{equation}
where $\pi$ represents joint probability distributions whose marginal distribution of $x$ is the assigned initial distribution and marginal distribution of $y$ is the assigned final distribution~\cite{peyre2019computational}. This quantity is the limit for $g\to0$ of the minimum entropy production in the overdamped dynamics~\eqref{eq:overdamped_lower_bound} multiplied by $t_f$~\cite{sanders2024optimal,AuGaMeMoMG2012}.

\section{Conclusions \& Outlook}
\label{sec:conclusions}
In this paper, we apply the multiscale perturbation expansion around the overdamped limit developed in~\cite{sanders2024optimal} to find minimal work protocols for inertial particles in non-harmonic traps. This approach allows us to make predictions for the optimal protocol and intermediate densities from its equivalent overdamped formulation, simplifying a complex numerical problem into a solution of the cell problem~\eqref{eq:cell}. This in turn can be solved 
using existing library implementations~\cite{flamary2021pot}.

Our approach allows considering transitions between non-Gaussian boundary conditions in the underdamped dynamics. This is demonstrated numerically with boundary conditions modeling nucleation, where we can make predictions for the behavior of the first and second order cumulants of the momentum process, as well as the shape of the position marginal probability density function. Agreement between the theoretical expectations and numerical simulations obtained by applying the prescribed protocol confirms the reliability of the expansion. 

Our numerical analysis allow us to detect qualitative differeneces between the considered underdamped regime and the corresponding overdamped one, both in the position marginal distribution (breaking of the symmetry $q-\av{q} \leftrightarrow \av{q}-q$) and in the control protocol. The momentum mean is shown to tend to a constant value during the evolution, while other quantities of interest follow a non-trivial path. We also use our method to compute the mean entropy production and verify the validity of the lower bounds provided in~\cite{sanders2024optimal}. Our numerical results support that inequality~\eqref{eq:overdamped_lower_bound} is tight.

Let us notice that the problem we considered is related to, but different from, what is typically done in the context of ``shortcuts to adiabaticity'', ``shortcuts to isothermality'', ``engineered swift equilibration'' and other techniques in control theory~\cite{MaPeGuTrCi2016, le2016fast, li2017shortcuts, guery2023driving}. While there, one is typically interested in finding a connection between two assigned end states, here we also require that connection to be optimal with respect to the energetic cost. The methods used to address the two problems are therefore quite different.

The effects of inertia and random noise cannot be ignored when designing electrical components at nanoscale. Understanding minimal work protocols for the erasure of one information bit in the underdamped dynamics is therefore important for improving the efficiency of computational devices, especially since these effects have been demonstrated to reduce the energy needed for bit operations~\cite{lopez2016sub,Deshpande2017}. Control of underdamped processes may have applications even in other branches of science: for instance, biological systems such as flocks of birds or self-propelled bacteria are known to be subject to inertial effects~\cite{Attanasi_2014, Scholz2018}. The application of control theory to such systems is a promising research line~\cite{Shankar_2022,baldovin2023, davis2023active}.

\FloatBarrier
\begin{acknowledgments}
 MB was supported by ERC Advanced Grant RG.BIO (Contract No. 785932). JS was supported by a University of Helsinki funded doctoral researcher position, Doctoral Program in Mathematics and Statistics. The authors wish to acknowledge CSC – IT Center for Science, Finland, for computational resources.
\end{acknowledgments}

\appendix

\section{Some remarks on notation}
\label{sec:appnotation}
The notation of this paper differs slightly from that of~\cite{sanders2024optimal}, where the physical observables are expressed as functions of multiple time scales, as required by the perturbative analysis. In this appendix, we provide a ``Rosetta stone'' to switch between the two notations. Quantities appearing in~\cite{sanders2024optimal} are marked with tildes. The time scales transform as
$$
\tilde{t}_0 = \frac{t}{\tau}\,, \quad \quad \tilde{t}_2 = \varepsilon^2 \frac{t}{\tau}\,.
$$
The fields solving the cell problem are related by
$$
 \tilde{\rho}_{\varepsilon^2 s}(x)=\rho(x,s)\,,\quad \quad  \tilde{\sigma}_{\varepsilon^2 s}(x)= \sigma(x,s)\,.
$$
Averages $\av{\cdot}$ in this paper are equivalent to $\tilde{\mathrm{E}}_{\mathcal{P}}\cbr{\cdot}$ there. For the position marginal distributions, one has
$$
\tilde{\mathrm{f}}_s(x) = \ell f(\ell x, \tau s)\,, \quad \tilde{\mathrm{f}}^{(0:2)}_{s,\varepsilon^2 s}(x) =  f_2(x, s)\,.
$$

\section{A symmetry conservation in the overdamped limit}
\label{app:symm}
We focus on the cell problem~\eqref{eq:cell}
with boundary conditions on $\rho(x,0)$ and $\rho(x,t_f)$. The problem is clearly equivalent to
\begin{subequations}
    \label{eq:cellalt}
    \begin{align}
        \partial_s \rho = \varepsilon^2 \partial_x (\rho\,\psi)\, \\
        \partial_s \psi = \varepsilon^2\psi\,\partial_x \psi\,,
    \end{align}
\end{subequations}
with $\psi=\partial_x \sigma$.
By defining $\mu(s)$ as in Eq.~\eqref{eq:mu}, it is possible to show that
$$
\ddot{\mu}=0\,.
$$
Indeed,
$$
\dot{\mu} = - \varepsilon^2 \int_{\mathbb{R}} dx \rho(x) \psi(x)
$$
so that
$$
\ddot{\mu}=-\varepsilon^4 \int_{\mathbb{R}} dx\, \psi \partial_x(\rho \psi)- \varepsilon^4 \int_{\mathbb{R}} dx\, \rho \psi \partial_x \psi = 0\,.
$$

Hence
\begin{equation}
  \mu(s)=\mu(0)+\dot{\mu} s  
\end{equation}
where $\dot{\mu}$ is a constant.

\begin{figure}
\centering    
\includegraphics[width=0.99\linewidth]{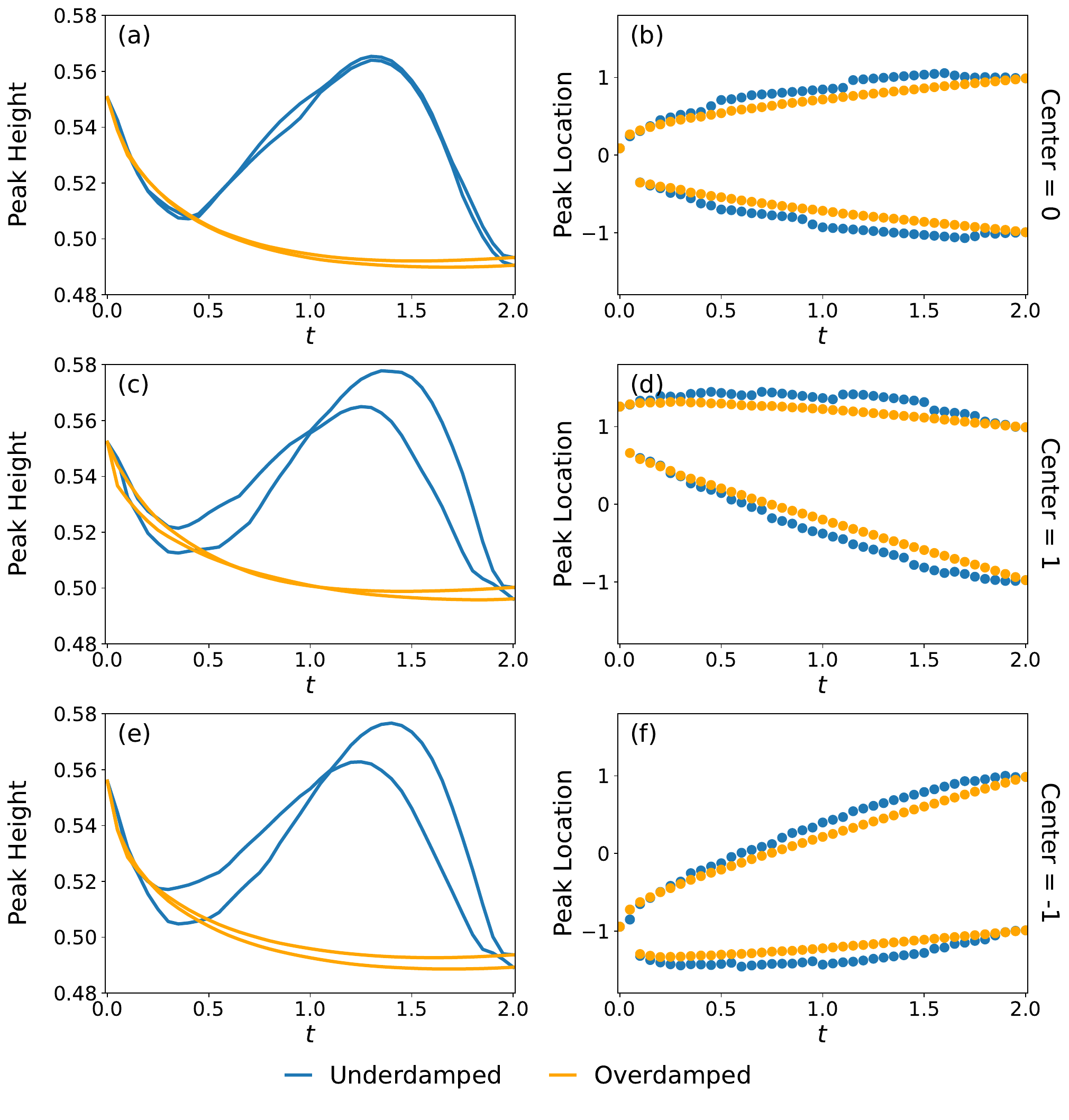}
    \caption{\label{fig:ep_momcumulants} Detail of symmetry breaking. Predictions for height (left) and location (right) of the peaks of the intermediate marginal probability density of the position of the nucelation process shown in Fig.~\ref{fig:potentials} in the underdamped (blue) and overdamped (orange) dynamics. We vary the center of the assigned initial distribution, $U_{\iota}(q) = (q-\alpha)^4$, with $\alpha=0$ in panels (a)-(b), $\alpha=1$ in panels (c)-(d), and  $\alpha=-1$ in panels (e)-(f). We compare the process in the overdamped (orange) and underdamped (blue) dynamics. Assigned final distribution and all other numerical parameters and method used are the same as in Fig.~\ref{fig:potentials}. Peaks are found using the signal processing library in scipy.}
\end{figure}

Let's consider the problem in the moving frame of $\mu(s)$, by defining the translated functions
\begin{equation}
\begin{aligned}
\wtr(x-\mu(s),s)&=\rho(x,s)\,\\
\wtp(x-\mu(s),s)&=\psi(x,s)\,.
\end{aligned}
\end{equation}
We want to show that if
\begin{equation}
\label{eq:symm}
\begin{aligned}
\wtr(x,0)&=\wtr(-x,0)\,\\
\wtr(x,t_f)&=\wtr(-x,t_f)\,,
\end{aligned}    
\end{equation}
then such symmetry is never broken along the dynamics.
To realize this, first notice that from Eq.~\eqref{eq:cellalt} one has
\begin{subequations}
    \label{eq:cell1}
    \begin{align}
    \partial_s \wtr - \dot{\mu} \,\partial_x \wtr&= \varepsilon^2 \partial_x( \wtr \wtp)\\
        \partial_s \wtp- \dot{\mu} \,\partial_x \wtp &= \varepsilon^2 \wtp \partial_x \wtp\,,
    \end{align}
\end{subequations}
or
\begin{subequations}
    \label{eq:cell2}
    \begin{align}
    \partial_s \wtr &= \varepsilon^2 \partial_x \cbr{\cbr{\wtp+ \frac{\dot{\mu}}{\varepsilon}}\wtr}\\
        \partial_s \wtp&= \varepsilon^2\cbr{\wtp+ \frac{\dot{\mu}}{\varepsilon}}\partial_x \wtp \,.
    \end{align}
\end{subequations}

But then, defining
\begin{equation}
    \varphi=\wtp+ \frac{\dot{\mu}}{\varepsilon}
\end{equation}
the problem can be recast into
\begin{subequations}
    \label{eq:cell3}
    \begin{align}
    \partial_s \wtr &= \varepsilon^2  \partial_x (\wtr \varphi)\\
        \partial_s \varphi&= \varepsilon^2 \varphi \partial_x \varphi\,,
    \end{align}
\end{subequations}
which has to be complemented with the boundary conditions for $\wtr$. This is exactly the same problem as~\eqref{eq:cellalt}. If $\varphi$ is odd at a given time, it will be odd forever. In this case $\wtr$ will be always even, if condition~\eqref{eq:symm} holds.

In Fig.~\ref{fig:ep_momcumulants}  we examine the behaviour of the peaks of the marginal density of the position in the underdamped dynamics. Using the predictions, we compute the heights and locations of the two peaks and compare these based on the location of the initial peak. If the boundary conditions are symmetric and they share the same value of the average position at $t=0$ and $t=t_f$, then the dynamics stays symmetric for the whole time interval both in the overdamped and in the underdamped case. If the boundary conditions are symmetric, but the average position changes between $t=0$ and $t=t_f$, then the overdamped dynamics stays symmetric, while the underdamped one breaks the position symmetry.

\bibliography{biblio}

\end{document}